\documentclass[11pt]{article}
\usepackage{jcappub}
\usepackage{bm}
\usepackage{color}
\usepackage{graphicx}
\usepackage{multirow}
\usepackage{enumitem}
\usepackage{cprotect}
\usepackage{amssymb}
\usepackage[normalem]{ulem}
\usepackage{fancyvrb}
\usepackage{upquote}
\usepackage{multirow}
\definecolor{ourcolor}{rgb}{0.7, 0.25, 0.05}

\newcommand{\qmsq}{\qmsq}
\renewcommand{\qmsq}{\q^2}

\newcommand{\be}{\begin{equation}}
\newcommand{\ee}{\end{equation}}
\newcommand{\een}{\end{subequations}}
\newcommand{\ben}{\begin{subequations}}
\newcommand{\beq}{\begin{eqalignno}}
\newcommand{\eeq}{\end{eqalignno}}

\newcommand{\lsim}{\mathrel{\mathop{\kern 0pt \rlap
      {\raise.2ex\hbox{$<$}}}\lower.9ex\hbox{\kern-.190em $ \sim$}}}
\newcommand{\gsim}{\mathrel{\mathop{\kern 0pt
      \rlap{\raise.2ex\hbox{$>$}}}\lower.9ex\hbox{\kern-.190em $\sim$}}}

\newcommand{\VectorTypefaceArrow}{
\let\oldvec\vec
\renewcommand{\vec}[1]{\oldvec{##1}} 
\newcommand{\uvec}[1]{\hat{##1}} 
}

\VectorTypefaceArrow

\usepackage{upgreek} 
\newcommand{\q}{{\widetilde{q}}}

\usepackage{mathtools}

\usepackage{booktabs}
\usepackage{xcolor}
\usepackage{arydshln}

\allowdisplaybreaks

\title{
Probing Non-Standard Neutrino Interactions with Interference: Insights from Dark Matter and Neutrino Experiments
}

\author[a, b]{Jong-Chul Park,}
\author[a]{Gaurav Tomar}
\affiliation[a]{Department of Physics and Institute of Quantum Systems (IQS), Chungnam National University, Daejeon 34134, Republic of Korea}
\affiliation[b]{Particle Theory and Cosmology Group, Center for Theoretical Physics of the Universe, Institute for Basic Science (IBS),  Daejeon, 34126, Republic of Korea}
\emailAdd{jcpark@cnu.ac.kr}
\emailAdd{tomarphysics@gmail.com}

\abstract{Neutrino-electron scattering experiments play a crucial role in investigating the non-standard interactions of neutrinos. 
In certain models, these interactions can include interference terms that may affect measurements. 
Next-generation direct detection experiments, designed primarily for dark-matter searches, are also getting sensitive to probe the neutrino properties. 
We utilise the data from XENONnT, a direct detection experiment, and Borexino, a low-energy solar neutrino experiment, to investigate the impact of interference on non-standard interactions. 
Our study considers models with an additional $U(1)$, including $U(1)_{B-L}$, $U(1)_{L_e-L_\mu}$, and $U(1)_{L_e-L_\tau}$, to investigate the impact of interference on non-standard neutrino interactions. We demonstrate that this interference can lead to a transition between the considered non-standard interaction models in the energy range relevant to both the XENONnT and Borexino experiments.
This transition can be used to distinguish among the considered models if any signals are observed at direct detection or neutrino experiments. 
Our findings underscore the importance of accounting for the interference and incorporating both direct detection and solar neutrino experiments to gain a better understanding of neutrino interactions and properties. 
}

\begin{document}

\maketitle

\section{Introduction}
\label{sec:introduction}

Currently, the investigation of neutrino properties is one of the most active fields of research in high-energy physics. 
Several historic milestones in neutrino physics such as supernova neutrino detection~\cite{Kamiokande-II:1987idp, Bionta:1987qt}, observations of neutrino oscillation~\cite{Super-Kamiokande:1998kpq, SNO:2002tuh, DayaBay:2012fng, RENO:2012mkc}, and measurements of coherent elastic neutrino-nucleus scattering~\cite{COHERENT:2017ipa, COHERENT:2020iec} provide us valuable information in understanding the Standard Model
(SM) of particle physics and the Universe. 
We can now describe how neutrinos interact with other SM particles based on the SM;  
however, the SM does not answer some fundamental questions about neutrinos~\cite{Bilenky:1987ty, Lesgourgues:2006nd, Avignone:2007fu}: e.g., neutrino mass generation and Majorana nature of neutrino. 
Therefore, neutrinos provide a definite pathway to search for the physics beyond the Standard Model.

In recent years, neutrino experiments~\cite{texono_2006, beda_2013, borexino_2000, borexino_2013, borexino2_2014, borexino_2017, coherent_2017, connie_2019, conus_2020, ccm_2021, colaresi_2022, pilar_2022} and dark-matter direct detection experiments~\cite{xenon_1t, xenonnt_resolution_2020, lz_22, xenonnt_2022, cdex_2022} have become increasingly important tools for studying the physics beyond the Standard Model~\cite{montanino_2008, harnik_2012, agarwalla_2012, khan_2019, amaral_2020, chen_2021, Jho:2020sku, Abdullah:2022zue, coloma_2022, khan_2022, dutta_2022, zeff_2022, rahul_2022}. 
Among these experiments, XENONnT~\cite{xenonnt_2022} and Borexino~\cite{borexino_2013, borexino2_2014, borexino_2017} are particularly noteworthy for their sensitivity to relatively low-energy new physics. 

The XENONnT~\cite{xenonnt_2022} experiment is located in the Gran Sasso underground laboratory and its primary aim is to detect weakly interacting massive particles which are one of the most popular candidates for dark matter. 
However, with its $\sim1$ keV energy threshold, XENONnT is also sensitive to elastic neutrino-electron scattering (E$\nu$ES) and can be used to study neutrino physics~\cite{rahul_2022}. 
The experiment uses a dual-phase time projection chamber filled with liquid xenon to detect the scintillation and ionization signals produced by particles interacting with the xenon nuclei. 
The XENONnT experiment has recently released its blind science results with an exposure of 1.16 tonne-years and achieved background reduction by a factor of five with respect to the XENON1T experiment~\cite{xenonnt_resolution_2020}. 
As a result, the well-appreciated XENON1T electron excess has been ruled out by the XENONnT experiment~\cite{xenonnt_2022}.

Borexino~\cite{borexino_2000, borexino_2013, borexino2_2014, borexino_2017}, on the other hand, is a low-energy solar neutrino experiment also located in the Gran Sasso laboratory. 
Its primary goal is to study the flux and energy spectrum of solar neutrinos, which can provide valuable information about the nuclear reactions that power the Sun. 
Borexino uses a large liquid scintillator detector to measure the signals produced by neutrino-electron scattering.

Non-standard interactions (NSIs) are extensions to SM that introduce new interactions between neutrinos and SM particles, with important implications for neutrino oscillation experiments~\cite{miranda_2015} and dark-matter direct detection experiments~\cite{cdex_2022, zeff_2022, rahul_2022}. 
Both XENONnT and Borexino experiments are sensitive to NSIs, which can modify the neutrino-electron scattering cross-sections. 
However, the experiments have not yet observed any signal, providing an opportunity to constrain NSI models using their data. 
A key feature of NSI models is the interference terms between the SM and NSI gauge bosons, which can significantly impact the experimental results~\cite{bilmis_2015}. 
Including interference effects is crucial when studying NSI models for XENONnT and Borexino experiments because they can lead to transitions between NSI models. 
Therefore, if a signal is observed in these experiments, interference effects can be used to identify the underlying NSI model. 
It is important to mention that without interference, there would be no transitions among NSI models in the regions of interest (ROI) for these experiments.

In our study, we focus on the NSI models for E$\nu$ES namely: $U(1)_{B-L}$, $U(1)_{L_e-L_\mu}$, and $U(1)_{L_e-L\tau}$, which produce constructive and destructive interference (clear from Eqs.~\ref{eq:zpcont}-\ref{eq:intcont} in coherence with Table~\ref{tab:gcharge}). 
We perform a detailed analysis of the behavior of these models in the context of Borexino Phase-II~\cite{borexino_2017} and XENONnT~\cite{xenonnt_2022} experiments. 
Our analysis show that for XENONnT the hierarchy among the NSI models persists even before and after including interference, while for Borexino a clear transition is observed among the different models. 
This transition among the NSI models can be particularly important to identify the specific NSI model for E$\nu$ES, that is responsible for a given signal in the experiments. 
In other words, if signals are observed at either the XENONnT or Borexino experiments, the interference effects could be used to determine which of the three NSI models is most likely responsible for the signals.

Overall, our study highlights the importance of considering the effects of interference when studying NSI models in the context of both neutrino and dark matter direct detection experiments. 
The ability to identify the specific NSI model responsible for a signal can provide valuable information for understanding the underlying physics that governs neutrino interactions with other particles.

The rest of the paper is organised as follows. 
In Section~\ref{sec:models}, we discuss the formalism of neutron-electron scattering.
In Section~\ref{sec:event_rate}, we describe the event rate calculation for the Borexino and XENONnT experiments. 
Finally, we provide our results in Section~\ref{sec:results} and summarize our work in Section~\ref{sec:summary}.

\section{Formalism}
\label{sec:models}

In this section, we examine the NSI models, exploring the interaction of neutrinos with electrons in the presence of a low-mass gauge boson. 
We closely follow the notations and formalism of Ref.~\cite{bhupal_2021} for our presentation. 
The general expression of the considered models in our analysis is expressed as following,
\begin{equation}
    - \mathcal{L} = g_V Q_{\ell} V_\mu \bar{\ell}\gamma^\mu \ell + g_V Q_{\nu_\ell} V_\mu \bar{\nu}_\ell\gamma^\mu \nu_\ell\,,
\end{equation}
where $g_V$ is the coupling of the light gauge boson $V$ and $Q_\ell$, $Q_{\nu_\ell}$ signify the gauge charges of SM leptons and neutrinos, respectively. 
The corresponding differential scattering cross-section
in the energy of the recoiling electron $E_e$ is easily available in the literature and quoted as follow,
\begin{equation}
    \frac{d\sigma_{\nu_l}}{dE_e}=\left[\frac{d\sigma_{\rm SM}}{dE_e}+\frac{d\sigma_V}{dE_e}+\frac{d\sigma_{\rm int}}{dE_e}\,\right]_{\nu_l}\,,
    \label{eq:ccs}
\end{equation}
where the first term includes the contribution from the $\nu_l$-$e$ scattering in the SM, while the second term corresponds to $\nu_l$-$e$ scattering in a NSI model. 
In the equation above, the interference between the SM gauge boson  and the dark gauge boson is encoded in the third term. 
The three terms are given by~\cite{bhupal_2021},
\begin{eqnarray}
    \frac{d\sigma_{\rm SM}}{dE_e} &=& \frac{2G_F^2m_e}{\pi E_\nu^2}\left [c_1^2E_\nu^2+c_2^2(E_\nu-E_e)^2-c_1c_2m_e E_e \right ]\,, \nonumber \\
    && \label{eq:smcont} \\
    \frac{d\sigma_V}{dE_e} &=&\frac{Q_{\nu_\ell}^2Q_e^2g_V^4m_e}{4\pi E_\nu^2}\frac{\left[2(E_\nu-E_e)E_\nu+(E_e-m_e)E_e \right]}{(2m_eE_e+m_V^2)^2}\,, \nonumber \\
    && \label{eq:zpcont}\\
    \frac{d\sigma_{\rm int}}{dE_e} &=& \frac{Q_{\nu_\ell}Q_eg_V^2G_Fm_e}{2\sqrt{2}E_\nu^2\pi(2m_eE_e+m_V^2)} 
    \times \left[c_3(2E_\nu^2-m_eE_e)+2c_4 (2E_\nu-E_e)E_e \right. \nonumber \\
    &+& \left. 4s_W^2\left\{2(E_\nu-E_e)E_\nu+(E_e-m_e)E_e\right\}\right]\,, \label{eq:intcont}
\end{eqnarray}
where $G_F$, $E_\nu$, $m_e$, and $m_V$ denote the Fermi constant, the incoming neutrino energy, the electron mass, and the $V$ gauge boson mass, respectively. 
The neutrino flavor-dependent coefficients, $c_1$ through $c_4$, are described in Table~\ref{tab:coeff}. 
In our analysis, we consider three NSI models, namely, $U(1)_{B-L}$, $U(1)_{L_e-L_\mu}$, and $U(1)_{L_e-L_\tau}$, which correspond to the gauge charges summarized in Table~\ref{tab:gcharge}.  
\begin{table}[t]
    \centering
    \begin{tabular}{c|c | c| c| c}
    \hline
         ~~~Flavor~~~ & ~~~~$c_1$~~~~ & ~~~~$c_2$~~~~ & ~~~~$c_3$~~~~ & ~~~~$c_4$~~~~  \\
         \hline 
         $\nu_e$ & $s_W^2+\frac{1}{2}$  & $s_W^2$ & $+1$ & $0$  
         \\
         $\nu_\mu,\nu_\tau$ & $s_W^2-\frac{1}{2}$ & $s_W^2$ & $-1$  & $0$ \\
         \hline 
    \end{tabular}
    \caption{A summary of the coefficients in Eqs.~\eqref{eq:smcont} through \eqref{eq:intcont}. 
    Here, $s_W$ represents the Weinberg angle $\theta_W$ as $s_W\equiv \sin\theta_W$.}
    \label{tab:coeff}
\end{table}
\begin{table}[t]
    \centering
    \begin{tabular}{c|c | c| c}
    \hline
         ~~~Charge/Model~~~ & ~~~~$U(1)_{B-L}$~~~~ & ~~~~$U(1)_{L_e-L_\mu}$~~~~ & ~~~~$U(1)_{L_e-L_\tau}$~~~~  \\
         \hline 
         $Q_{\nu_e}$ & $-1$  & $+1$ & +1  
         \\
         $Q_{\nu_\mu}$ & $-1$  & $-1$ & $0$     \\
         $Q_{\nu_\tau}$ & $-1$  & $0$ & $-1$    \\
         \hline 
    \end{tabular}
    \caption{A summary of the gauge charges of the models considered in our analysis.}
    \label{tab:gcharge}
\end{table}

\section{Event rate for the Borexino and XENONnT experiments}
\label{sec:event_rate}
Neutrino scattering experiments can be utilised to search for light dark gauge boson. 
It is clear from Eqs.~\ref{eq:zpcont}--\ref{eq:intcont} that the differential cross-section increases for low recoil energy of electron and for a small value of $M_V$. 
As a result, low threshold neutrino experiments are ideal place to search for the interference effects. 
Borexino~\cite{borexino_2000} is such an experiment we consider in our analysis, that  measures the solar neutrinos~\cite{borexino_2008, borexino_2017}. 
Specifically, we utilise the spectrum of $^7$Be monochromtic solar neutrino (with 0.862 MeV energy)  measured by elastic scattering of neutrinos with liquid scintillator (doped pseudocumene). 
The chemical composition of pseudocumene is C$_9$H$_{12}$ with atomic mass 120.19 and 66 electrons per molecule. 
As a result the number of electrons in 100 tons of pseudocumene are $N_e=3.307\times 10^{31}$. 
Specifically, for our $^7$Be solar neutrino analysis, we utilise the Borexino phase-II data~\footnote{While we have analyzed only the $^7$Be Borexino phase-II data, we want to highlight that the main conclusions of our study will remain unchanged even if data from other solar species were excluded.} with $48.3\pm 1.1$ cpd/100 tons events~\cite{borexino_2017} and high-metallicity standard solar model flux, $\Phi_{^7\rm Be}=4.93\times 10^{9}$ cm$^{-2}$s$^{-1}$~\cite{end_point}.

The dark matter direct detection experiments have reached an unprecedented sensitivities and now are capable to explore low energy neutrino physics. 
In our analysis, we consider XENONnT experiment~\cite{xenonnt_2022} with 1 keV experimental threshold which is two order of magnitude better than dedicated neutrino experiments. 
With an exposure of 5.9 tons, its ROI spreads between 1-30 keV. 
XENONnT is a dual-phase detector, looking for scintillation photons and subsequent ionisation electrons through $S_1$ and $S_2$ signals respectively, in the aftermath of a dark matter-nucleus scattering or background events. 
The main background in the electron recoil spectrum of XENONnT comes from $\beta$-decay events and E$\nu$ES which mainly gets contributions from $pp$ and $^7$Be (with 0.862 MeV energy) components of the solar neutrino spectrum~\cite{end_point}. 
The other components contribute negligibly to the solar neutrino spectrum.

The differential E$\nu$ES rate in considered experiments is calculated as following,
\begin{equation}
 \frac{dR}{dE_e} =  N_T \sum_{i} \int^{E^{\rm max}_\nu} _{E^{\rm min}_\nu} dE_\nu \frac{d\Phi^{\nu}_i(E_\nu)}{E_\nu}\left(\frac{d\sigma_{\nu e}}{dE_e} P_{ee}+\sum_{f=\mu, \tau} \frac{d\sigma_{\nu f}}{dE_e} P_{ef}
 \right)\,,
 \label{eq:diff_rate}
\end{equation}
where the differential cross-sections (Eqs.~\ref{eq:smcont}-\ref{eq:intcont}) are weighted by the survival probabilities, arising due to neutrino propagation from the Sun to the Earth, resulting neutrino oscillations. 
In the equation above, $P_{ee}={\rm cos}^4 \theta_{13} P_{eff}+ {\rm sin}^4 \theta_{13} $ is the survival probability of the solar neutrino, calculated in the two-flavour approximation, $P_{e\mu}=(1-P_{ee}){\rm cos}^2\theta_{23}$, and $P_{e\tau}=(1-P_{ee}){\rm sin}^2\theta_{23}$.
Here, $P_{eff}=(1+{\rm cos2\theta_M}\rm{cos}2\theta_{12})$ which depends also on the neutrino propagation path and accounts for matter effects~\cite{parke_1986}. 
$P_{eff}$ is evaluated by taking the current best-fit values of the oscillation parameters~\cite{fit_oscillation_2020} and assuming their central values for normal ordering.

In the Eq.~\ref{eq:diff_rate}, $N_T$ represents the number of target electrons; for the Borexino experiment, $N_T=3.307\times 10^{31}$ is the number of targets per 100 tons, while for the XENONnT experiment, $N_T=Z_{eff}(E_e)m_{\rm det}N_A/m_{\rm Xe}$ where $m_{\rm Xe}$ is the molar mass of xenon, $m_{\rm det}$ is the fiducial mass of the detector, and $N_A$ is the Avogadro number. 
Here, $Z_{eff} (E_e)$ represents the number of electrons that can be ionised by certain energy deposition $E_e$.  Following~\cite{zeff_2022}, we approximated the latter through a series of steps functions. 
An alternative method is the relativistic random-phase approximation theory which improves the calculation by utilising the atomic many body effects~\cite{rrpa_2016}. 
Moreover, $\Phi_i^{\nu}$ represents the solar neutrino flux, which is dominated by $pp$ and $^7$Be (with 0.862 MeV energy) neutrinos in the ROI explored by the XENONnT experiment. 
While for the Borexino experiment, the line spectrum of $^7$Be is considered for $\Phi_i^{\nu}$. 
The minimum neutrino energy required to register the electron recoil $E_e$ is given by
\begin{equation}
  E^{\rm min}_\nu=\frac{E_e+\sqrt{E^2_e+2m_e E_e}}{2}\,,
\end{equation}
while the maximum neutrino energy $E_{\nu}^{\rm max}\sim 20$ MeV~\cite{end_point}. 
The resolution and efficiency for the XENONnT experiment are 
taken from~\cite{xenonnt_resolution_2020} and~\cite{xenonnt_2022}, respectively.  
In the following section, we will describe our results by performing a chi-square fit to the observed data with our benchmark models.

\section{Results}
\label{sec:results}

We constrain the new physics parameters by employing a Gaussian $\chi^2$ function,
\begin{equation}
 \chi^2(\mathcal{S})=\sum_{k}\left(\frac{R^k_{\rm exp}-R^k_{{\rm SM}+V+B_0}(\mathcal{S}, \beta)}{\sigma_k}\right)^2+\left(\frac{\beta^k}{\sigma_\beta^k}\right)^2\,,
 \label{eq:chi2_pen}
 \end{equation}
where $k$ runs over the energy bins, $\sigma_k$ refers to statistical uncertainty per bin for a considered experiment. 
We consider 30 bins for our XENONnT analysis similar to the published result~\cite{xenonnt_2022}, while for Borexino we only consider $^7$Be (with 0.862 MeV energy) phase-II data in an extended energy range between 0.19-2.93 MeV~\cite{borexino_2017}. 
In the equation above, $R^k_{\rm exp}$ stands for the experimental differential rate (event rate) for the XENONnT (Borexino) experiment, and $R^k_{{\rm SM}+V+B_0} (\mathcal{S}, \beta)=(1+\beta)R^k_{{\rm SM}+V}$+$B^k_0$ where $B^k_0$ is the background reported by the relevant experiment. Here, $R^k_{{\rm SM}+V}$ includes the contributions from the SM and dark gauge boson, which are calculated using Eq.~\ref{eq:diff_rate} for the corresponding energy bin.
In our XENONnT analysis $B^k_0$ is obtained from Ref.~\cite{xenonnt_2022} after subtracting the SM E$\nu$ES contribution, while for the Borexino experiment background is already subtracted and so we do not include any background terms in our $\chi^2$ model. 
Moreover, $\beta^k$ in the above equation is the penalty factor which we utilise to incorporate the uncertainty in the solar neutrino flux. 
In our analysis, we consider $\sigma_\beta^k=7\%$~\cite{flux_uncertainity, zeff_2022}.
The bound on the new physics parameters is obtained by,
\begin{equation}
 \chi^2-\chi^2_{\rm min}=2.71\,,
 \label{eq:chi2_min}
\end{equation}
which corresponds to $90\%$ C.L. sensitivity reach for one degrees of freedom. 
We obtain our results with Eq.~\ref{eq:chi2_min} for $U(1)_{B-L}, U(1)_{L_e-L_\mu}$, and $U(1)_{L_e-L_\tau}$ models, while utilising the XENONnT and Borexino experiments.

\begin{figure}[t]
\begin{center}
    \includegraphics[width=14 cm]{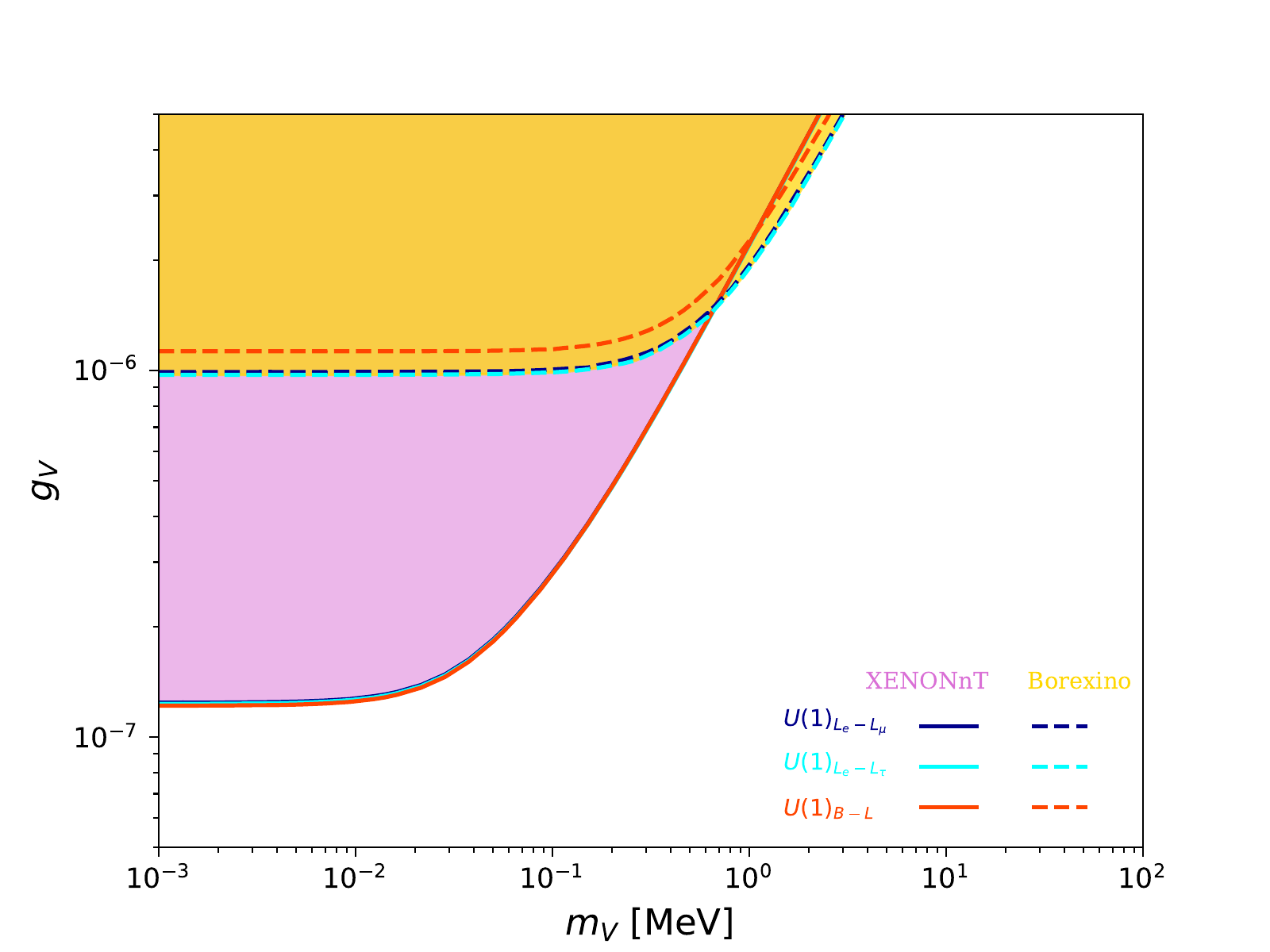}
    \end{center}
  \caption{The exclusion limits for the $U(1)_{B-L}$, $U(1)_{L_e-L_\mu}$ and $U(1)_{L_e-L_\tau}$ models, including the interference between the SM and the dark sector, for both the XENONnT (solid lines) and Borexino (dashed lines) experiments.}
  \label{fig:xenonnt_models}
\end{figure}

In Fig.~\ref{fig:xenonnt_models}, we show the constrained regions by Borexino (dashed lines) and XENONnT (solid lines) in the $(m_V, g_V)$ plain at 90$\%$ C.L. for $U(1)_{B-L}$ (red lines), $U(1)_{L_e-L_\mu}$ (blue lines), and $U(1)_{L_e-L_\tau}$ (cyan lines) models.
While obtaining the exclusion limits, we include the interference terms, due to which a hierarchy among the considered models is observed.  
Clearly, XENONnT~\cite{xenonnt_2022} puts more stringent limits on the most of parameter space of the considered NSI models, surpassing the Borexino phase-II data~\cite{borexino_2017}. 
The jiangmen underground neutrino observatory (JUNO) experiment is proposed to have a large fiducial mass of around 20 kton and excellent energy resolution of $0.03/ \sqrt{E(\rm MeV)}$~\cite{juno_2015}. 
This makes it capable of enhancing the sensitivity of Borexino by up to an order of magnitude. 
It is also important to note that the upcoming installation of the liquid scintillator-based detector (LSC) at Yamilab~\cite{Seo:2020dtx, lsc_2022} will bring significant improvements to solar neutrino measurements compared to the current Borexino experiment, which can therefore provide better sensitivities for NSI model searches.

\begin{figure}[t]
    \includegraphics[width=8.5 cm]{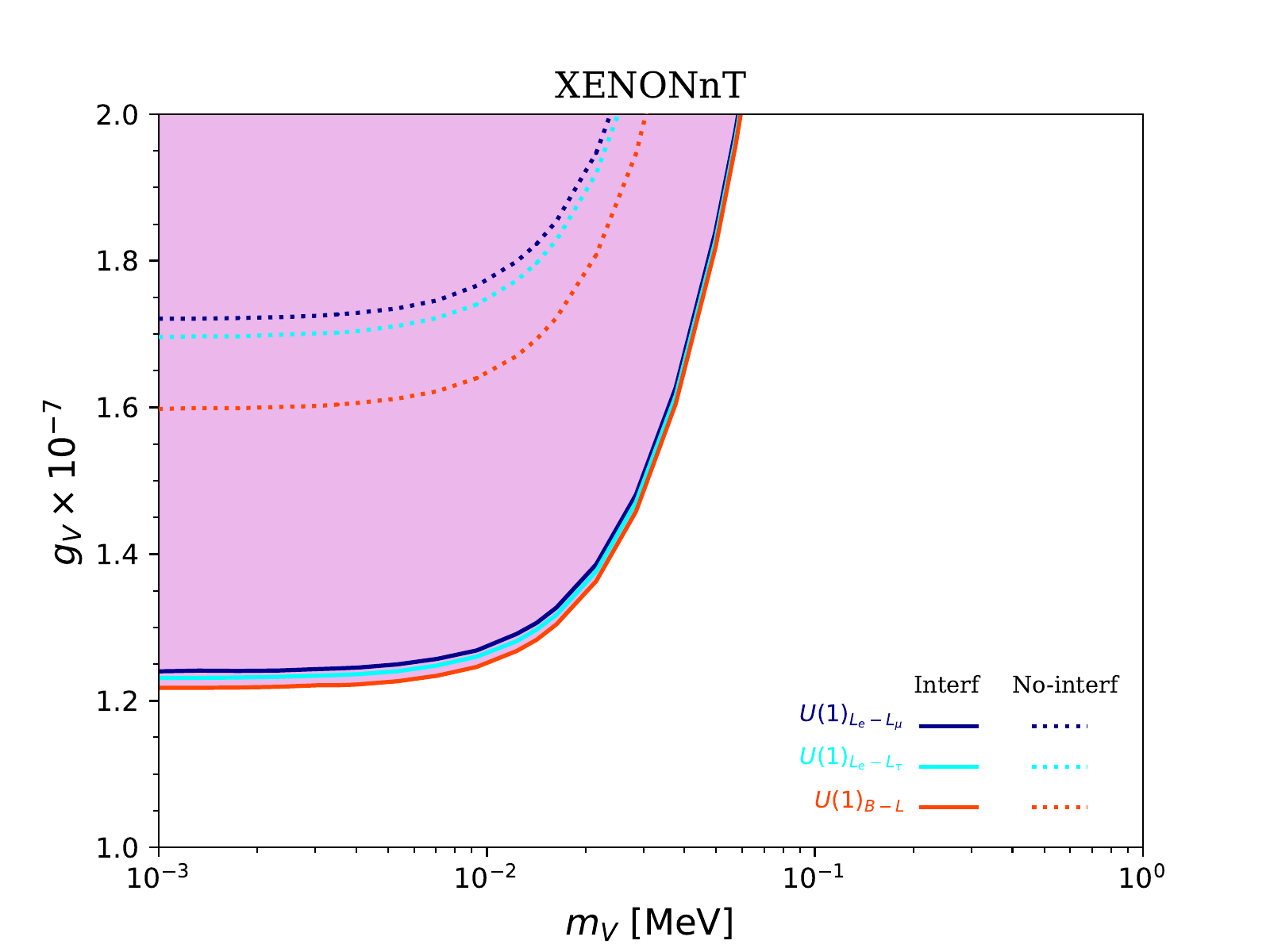}
    \includegraphics[width=8.5 cm]{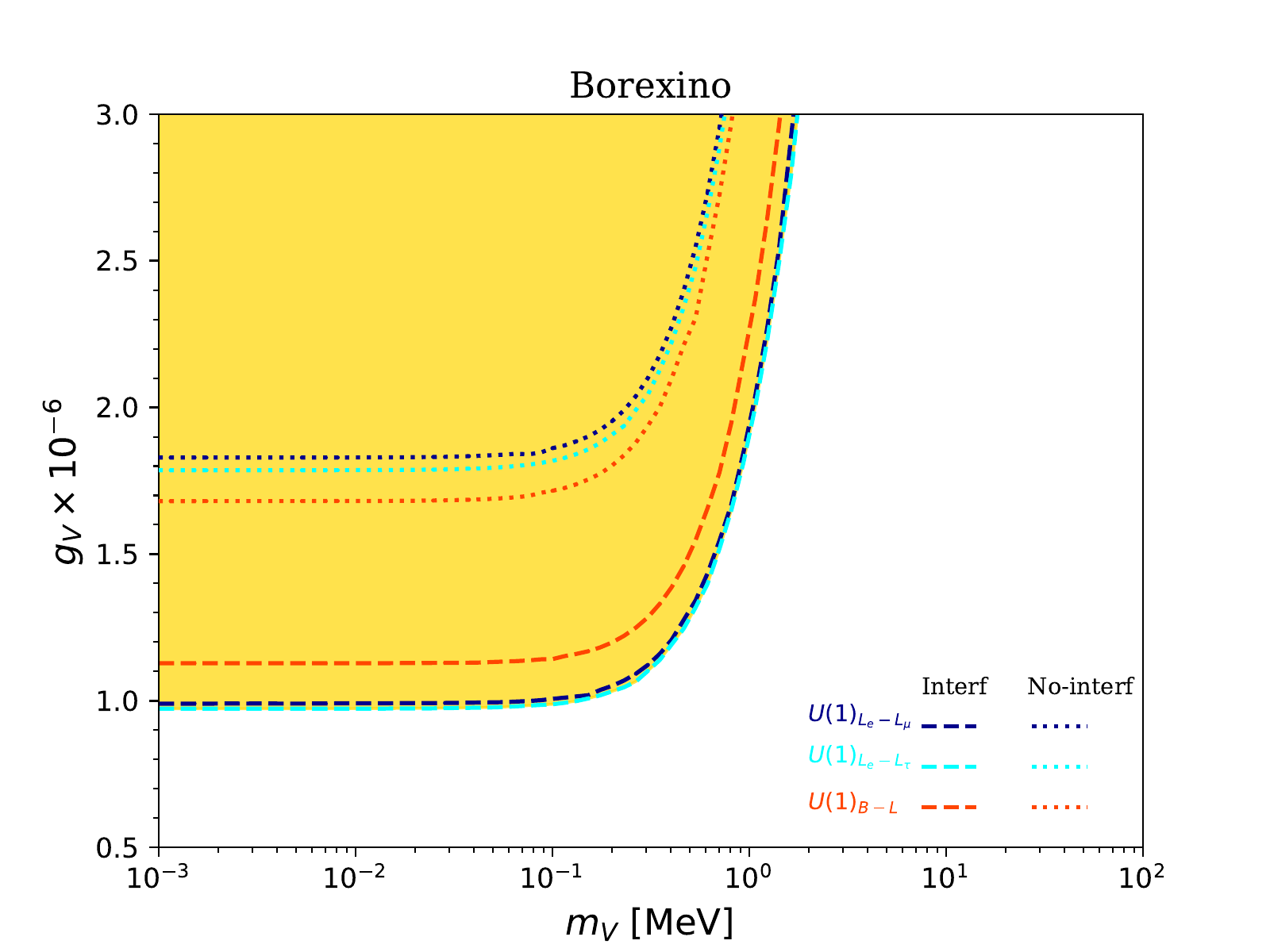}

  \caption{A zoomed version of Fig.~\ref{fig:xenonnt_models} for the XENONnT (left) and Borexino (right) experiments. 
  The solid lines represent the limits including interference, while the dotted lines show the limits excluding interference for XENONnT. 
  For Borexino, the limits with interference are presented by the dashed lines, while the dotted lines represent the case of no-interference.}
  \label{fig:xenonnt_models_zoomed}
\end{figure}

A zoomed version of Fig.~\ref{fig:xenonnt_models} is depicted in Fig.~\ref{fig:xenonnt_models_zoomed}, where the difference among NSI models is clearly appreciated for the XENONnT and Borexino experiments. 
For brevity, we presented exclusion limits with interference (solid lines for XENONnT and dashed lines for Borexino) and without interference (dotted lines) for the considered models. 
It is puzzling that in the Borexino experiment without including interference term $U(1)_{B-L}$ is the most constraining model, while inclusion of interference term leads to $U(1)_{L_e-L_\tau}$ as the most constraining model. 
The situation is different for the XENONnT experiment where $U(1)_{B-L}$ is always the most constraining model irrespective of interference between the SM and dark photon interaction.

\begin{figure}[t]
    \includegraphics[width=8.5cm]{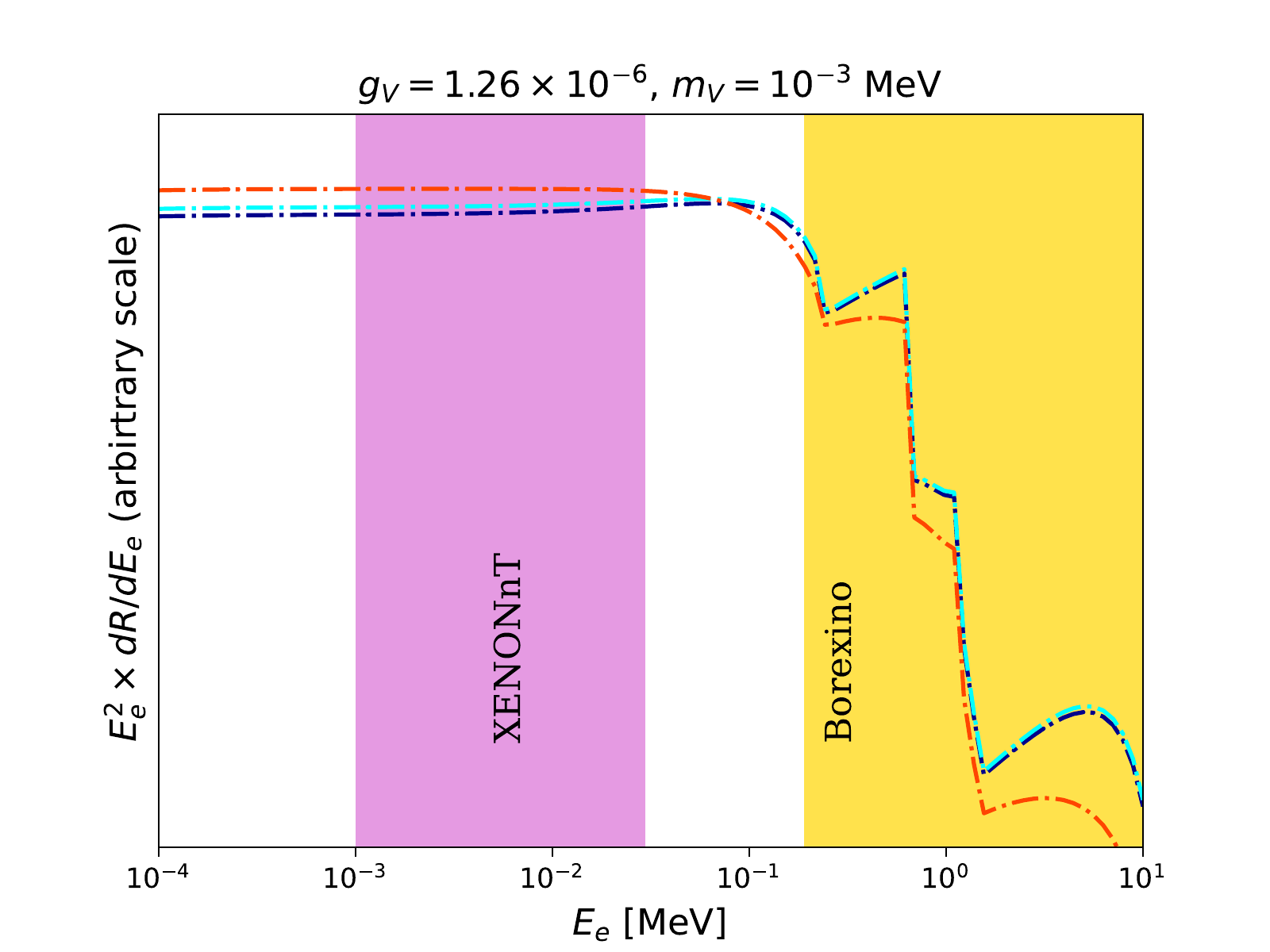}
    \includegraphics[width=8.5cm]{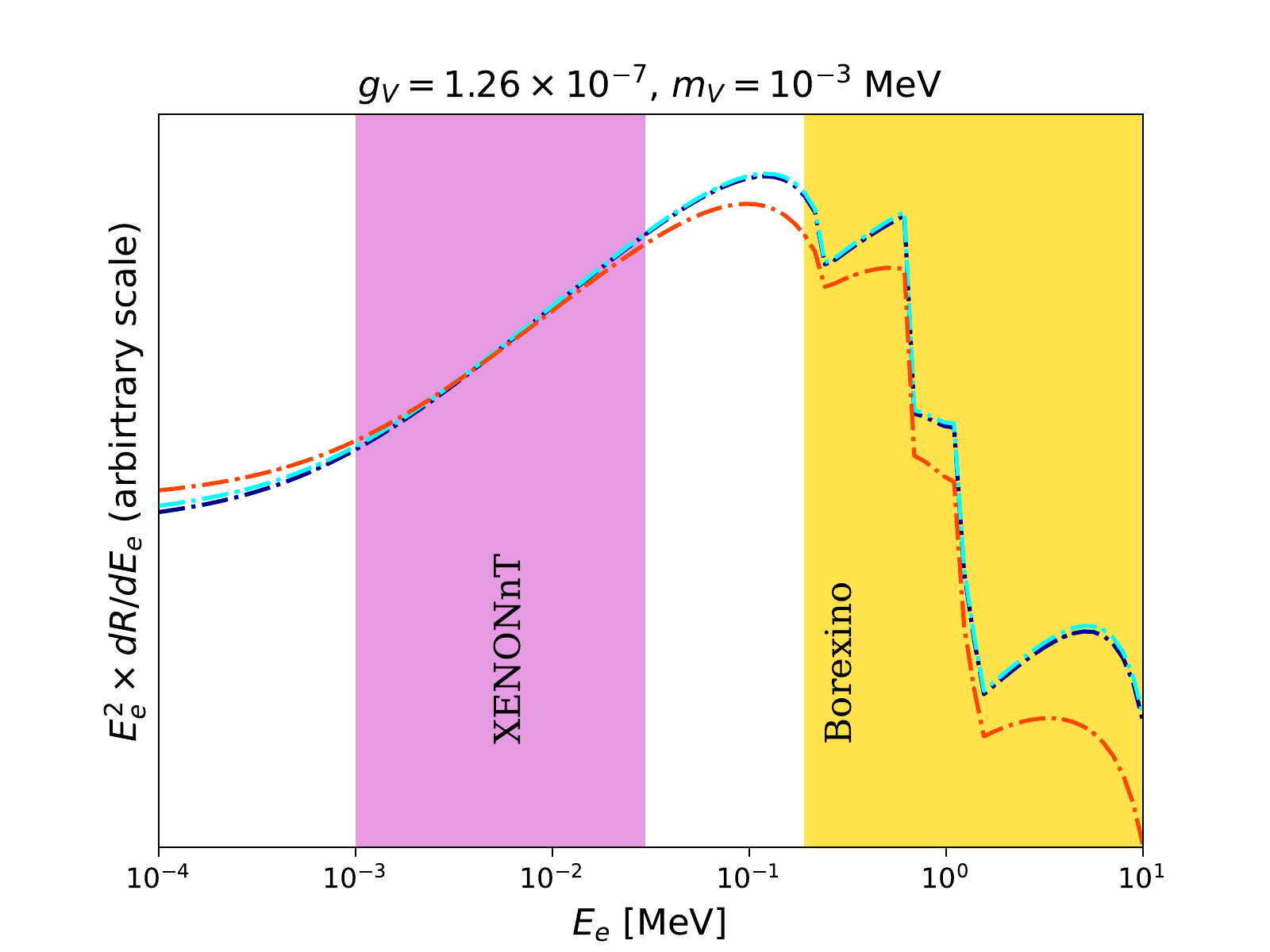}\\

    \includegraphics[width=8.5cm]{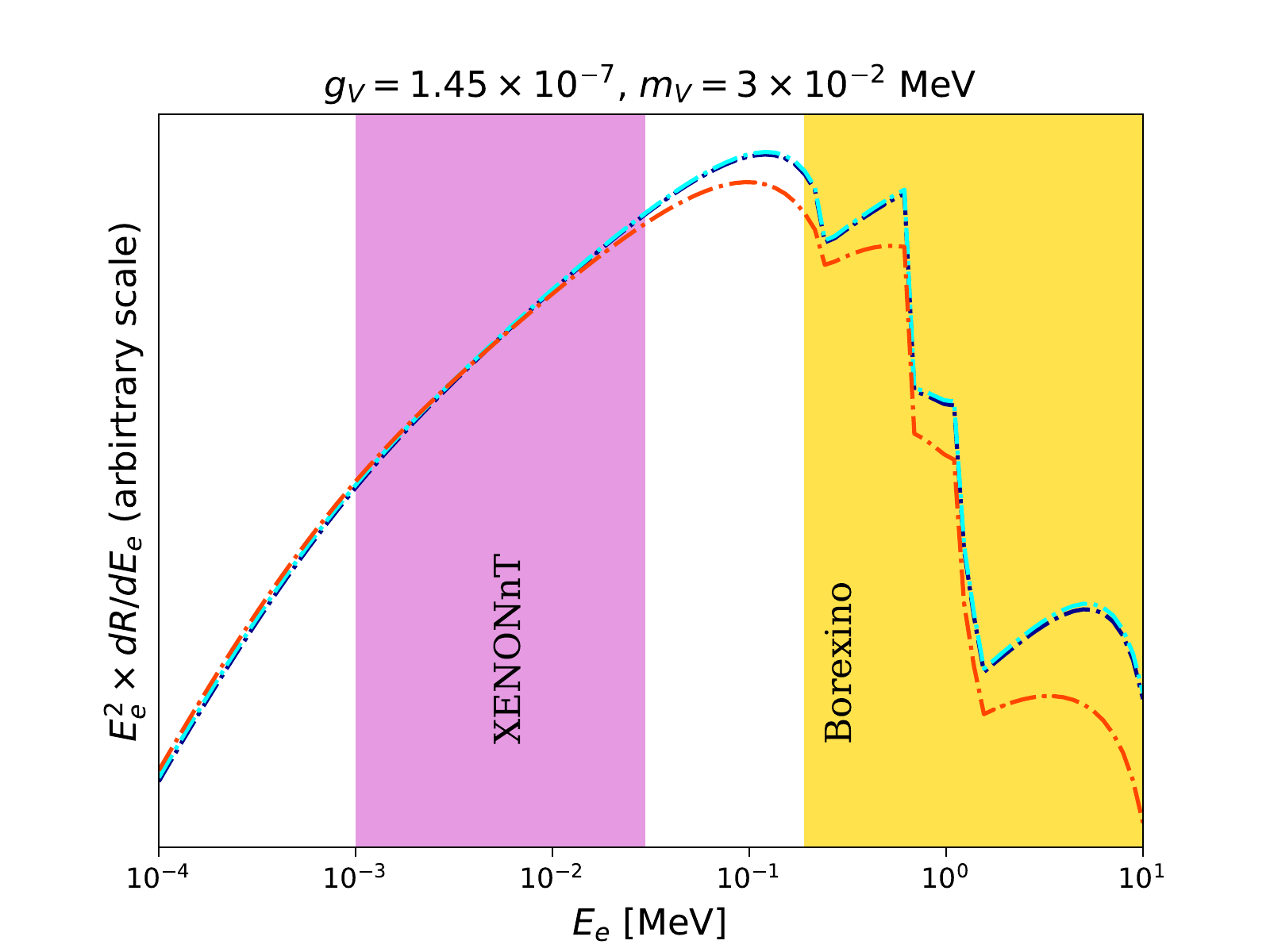}
    \includegraphics[width=8.5cm]{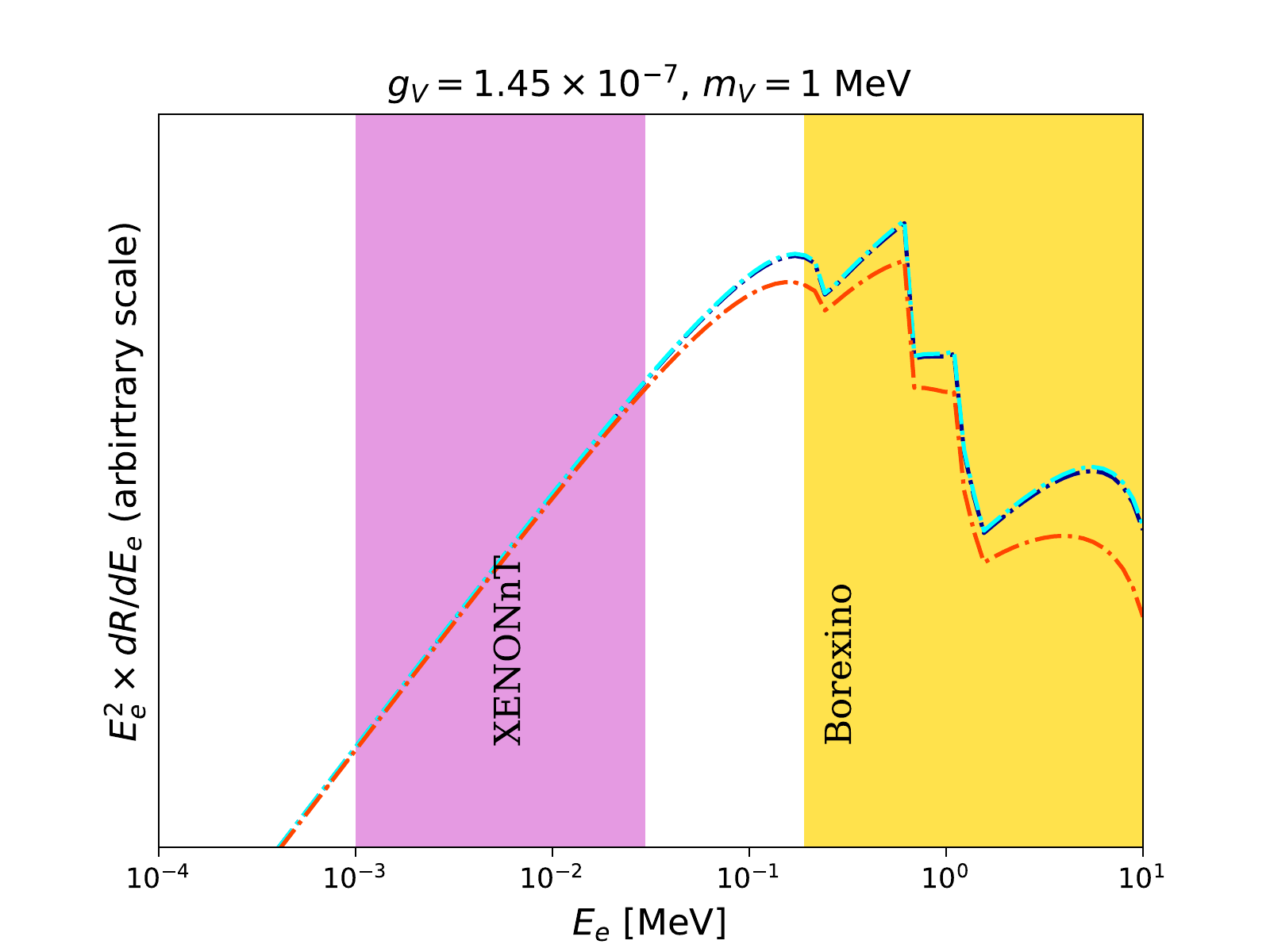}
  \caption{The differential rates in the $U(1)_{B-L}$ (red), $U_{L_e-L_\mu}$ (dark blue) and $U_{L_e-L_\tau}$ (cyan) models for the energy range relevant for Borexino and XENONnT experiments.}
  \label{fig:diff_rate}
\end{figure}

\begin{figure}
\begin{center}
    \includegraphics[width=14cm]{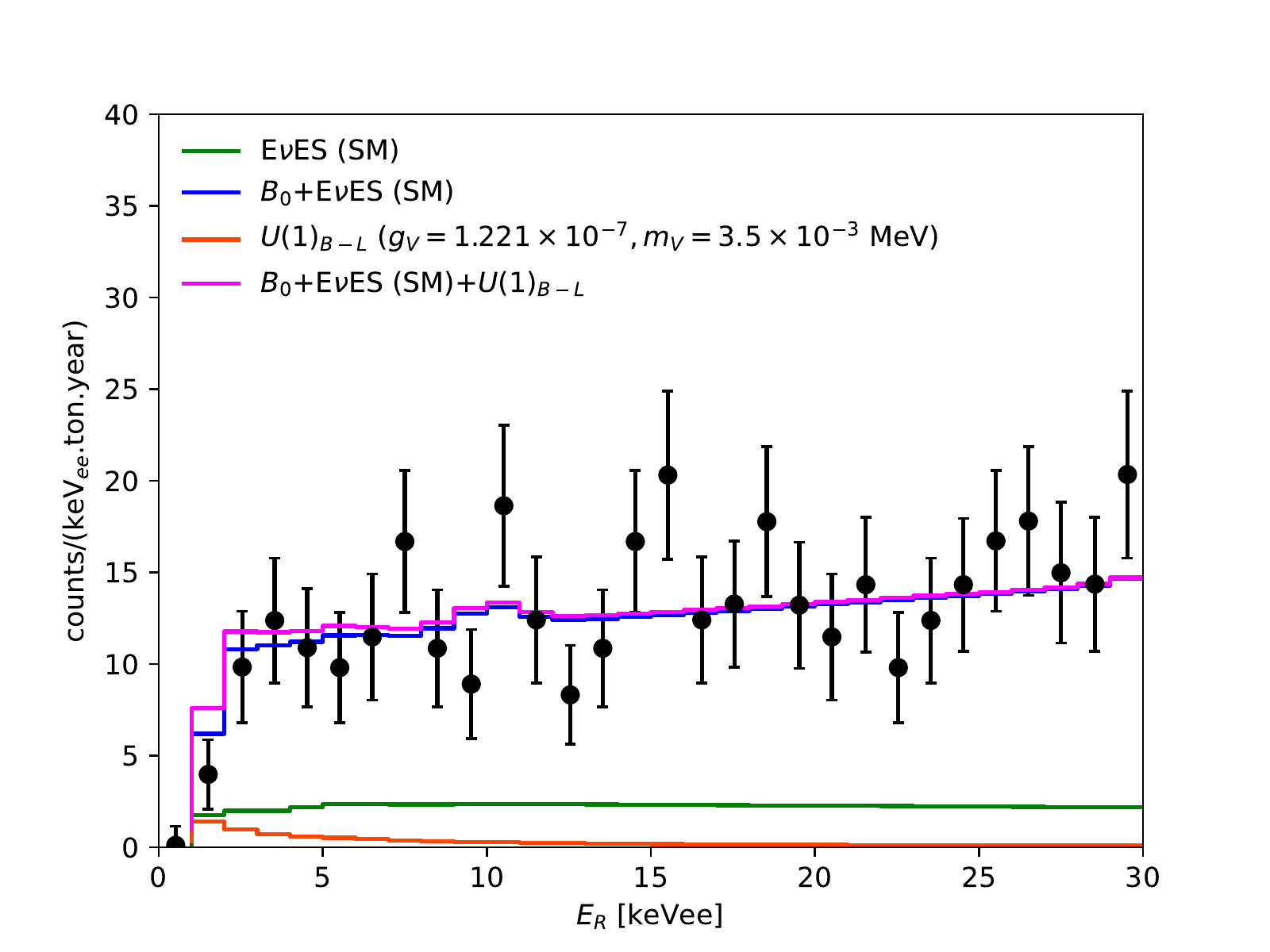}
    \end{center}
  \caption{Expected signal from the $U(1)_{B-L}$ model and comparison with the experimental data at XENONnT~\cite{xenonnt_2022}. 
  Here, the black dots with the error-bar represents the XENONnT data, while $B_0$ is the background obtained from Ref.~\cite{xenonnt_2022} after subtracting the SM E$\nu$ES contribution.
  }
  \label{fig:xenonnt_bml}
\end{figure}

The solution to this conundrum is discovered in Fig.~\ref{fig:diff_rate} where the differential rate for the considered models is shown as a function of electron recoil energy $E_e$.  
For clarity of transition behavior, we have plotted $E^2_e \times dR/dE_e$ instead of $dR/dE_e$ in Fig.~\ref{fig:diff_rate}. 
The shaded purple region represents the ROI for the XENONnT experiment while the shaded gold region starts from 0.190 MeV, relevant for the Borexino experiment. 
There are presented the results for four representative cases in variation of $g_V$ and $m_V$, for which the transition behavior can be understood utilising Eqs.~\ref{eq:zpcont}-\ref{eq:intcont}. 
In the case of a large coupling $g_V$ and small mediator mass $m_V$, the dark photon term remains competitive with the interference term, showing a clear transition. 
As $g_V$ decrease, keeping small $m_V$, the transition still takes place but not as prominent. 
In the case of a heavy mediator and small coupling, the contribution from interference term dominates over the dark photon contribution. 
As a result there is no clearly visible transition for the considered model. 
We find that because of the interference term $U(1)_{B-L}$ is the most constraining model in the ROI of XENONnT,\footnote{At some of the points of the parameter space, transition occurs within the ROI of XENONnT, but is not observed in the exclusion limits. 
It has been found that the $U(1)_{B-L}$ model is consistently the most restrictive in the exclusion limits due to the dominant contribution of the initial energy bins (See Fig.~\ref{fig:xenonnt_bml}).} 
while $U(1)_{L_e-L_\tau}$ dominates in the region concerned to Borexino. 
This highlights the importance of considering the Borexino and XENONnT experiment with complimentary ROIs in our analysis. 
By utilising their data, it would be possible to distinguish different models.

It is important to note that regardless of the model considered, the inclusion of interference terms strengthens the exclusion limits, indicating the presence of constructive interference. 
In particular, the $U(1)_{B-L}$ model exhibits constructive interference for all neutrino flavors, whereas the $U(1)_{L_e-L_\mu}$ and $U(1)_{L_e-L_\tau}$ models can exhibit both constructive and destructive interference, depending on the neutrino flavor.

This phenomenon is explained by the weighted combinations of cross-sections, as described in Eq.~\ref{eq:diff_rate}. 
In the $U(1)_{B-L}$ model, the same charge ($Q_{\nu_l}=Q_e=-1$) for all neutrino flavors leads to constructive interference. 
In contrast, the $U(1)_{L_e-L_\mu}$ model has charges of $Q_{\nu_e}=Q_e=1$, $Q_{\nu_\mu}=-1$, and $Q_{\nu_\tau}=0$ while the $U(1)_{L_e-L_\tau}$ model has charges of $Q_{\nu_e}=Q_e=1$, $Q_{\nu_\mu}=0$, and $Q_{\nu_\tau}=-1$. 
As a result, $\nu_e$ experiences constructive interference, while $\nu_\mu$ or $\nu_\tau$ experience destructive interference in these models. 
Despite these variations, the dominance of the electron-neutrino scattering cross-section over the muon-neutrino and tau-neutrino scattering cross-sections leads to an overall constructive interference between the SM gauge boson and the dark gauge boson across all models.

In Fig.~\ref{fig:xenonnt_bml}, we show the expected signal from the $U(1)_{B-L}$ model
with the background ($B_0$) obtained from Ref.~\cite{xenonnt_2022} and the SM E$\nu$ES contribution, which is compared with the observed data by the XENONnT experiment~\cite{xenonnt_2022}. 
The chosen benchmark is consistent with Fig.~\ref{fig:xenonnt_models}.

\section{Summary}
\label{sec:summary}
We studied the impact of the interference between the standard and non-standard neutrino interactions in $U(1)$-extended models while analysing the data from the XENONnT and Borexino
experiments. 
Our study underlines a transition between different non-standard interaction models because of the presence of the interference between the standard and non-standard interactions. 
Depending upon the vector-mediator coupling to the SM and its mass, the hierarchy among considered models is opposite in nature for the ROIs of the XENONnT and Borexino experiments. 
In the epoch of observed signals at direct detection or neutrino experiments, this transition can be used to distinguish among considered NSI models. 
Basically, our findings underscore the importance of incorporating both direct detection and solar neutrino experiments to gain a better understanding of neutrino interactions and properties.

\acknowledgments 
We would like to thank Monojit Ghosh for useful discussions.
The work is supported by the National Research Foundation of Korea (NRF) [NRF-2019R1C1C1005073].
The work of JCP is partially supported by IBS under the project code, IBS-R018-D1.

\providecommand{\href}[2]{#2}\begingroup\raggedright\endgroup


\begin{thebibliography}{10}

\bibitem{Kamiokande-II:1987idp}
{\scshape Kamiokande-II} collaboration, K.~Hirata et~al., \emph{{Observation of
  a Neutrino Burst from the Supernova SN 1987a}},
  \href{https://doi.org/10.1103/PhysRevLett.58.1490}{\emph{Phys. Rev. Lett.}
  {\bfseries 58} (1987) 1490--1493}.

\bibitem{Bionta:1987qt}
R.~M. Bionta et~al., \emph{{Observation of a Neutrino Burst in Coincidence with
  Supernova SN 1987a in the Large Magellanic Cloud}},
  \href{https://doi.org/10.1103/PhysRevLett.58.1494}{\emph{Phys. Rev. Lett.}
  {\bfseries 58} (1987) 1494}.

\bibitem{Super-Kamiokande:1998kpq}
{\scshape Super-Kamiokande} collaboration, Y.~Fukuda et~al., \emph{{Evidence
  for oscillation of atmospheric neutrinos}},
  \href{https://doi.org/10.1103/PhysRevLett.81.1562}{\emph{Phys. Rev. Lett.}
  {\bfseries 81} (1998) 1562--1567},
  [\href{https://arxiv.org/abs/hep-ex/9807003}{{\ttfamily hep-ex/9807003}}].

\bibitem{SNO:2002tuh}
{\scshape SNO} collaboration, Q.~R. Ahmad et~al., \emph{{Direct evidence for
  neutrino flavor transformation from neutral current interactions in the
  Sudbury Neutrino Observatory}},
  \href{https://doi.org/10.1103/PhysRevLett.89.011301}{\emph{Phys. Rev. Lett.}
  {\bfseries 89} (2002) 011301},
  [\href{https://arxiv.org/abs/nucl-ex/0204008}{{\ttfamily nucl-ex/0204008}}].

\bibitem{DayaBay:2012fng}
{\scshape Daya Bay} collaboration, F.~P. An et~al., \emph{{Observation of
  electron-antineutrino disappearance at Daya Bay}},
  \href{https://doi.org/10.1103/PhysRevLett.108.171803}{\emph{Phys. Rev. Lett.}
  {\bfseries 108} (2012) 171803},
  [\href{https://arxiv.org/abs/1203.1669}{{\ttfamily 1203.1669}}].

\bibitem{RENO:2012mkc}
{\scshape RENO} collaboration, J.~K. Ahn et~al., \emph{{Observation of Reactor
  Electron Antineutrino Disappearance in the RENO Experiment}},
  \href{https://doi.org/10.1103/PhysRevLett.108.191802}{\emph{Phys. Rev. Lett.}
  {\bfseries 108} (2012) 191802},
  [\href{https://arxiv.org/abs/1204.0626}{{\ttfamily 1204.0626}}].

\bibitem{COHERENT:2017ipa}
{\scshape COHERENT} collaboration, D.~Akimov et~al., \emph{{Observation of
  Coherent Elastic Neutrino-Nucleus Scattering}},
  \href{https://doi.org/10.1126/science.aao0990}{\emph{Science} {\bfseries 357}
  (2017) 1123--1126}, [\href{https://arxiv.org/abs/1708.01294}{{\ttfamily
  1708.01294}}].

\bibitem{COHERENT:2020iec}
{\scshape COHERENT} collaboration, D.~Akimov et~al., \emph{{First Measurement
  of Coherent Elastic Neutrino-Nucleus Scattering on Argon}},
  \href{https://doi.org/10.1103/PhysRevLett.126.012002}{\emph{Phys. Rev. Lett.}
  {\bfseries 126} (2021) 012002},
  [\href{https://arxiv.org/abs/2003.10630}{{\ttfamily 2003.10630}}].

\bibitem{Bilenky:1987ty}
S.~M. Bilenky and S.~T. Petcov, \emph{{Massive Neutrinos and Neutrino
  Oscillations}}, \href{https://doi.org/10.1103/RevModPhys.59.671}{\emph{Rev.
  Mod. Phys.} {\bfseries 59} (1987) 671}.

\bibitem{Lesgourgues:2006nd}
J.~Lesgourgues and S.~Pastor, \emph{{Massive neutrinos and cosmology}},
  \href{https://doi.org/10.1016/j.physrep.2006.04.001}{\emph{Phys. Rept.}
  {\bfseries 429} (2006) 307--379},
  [\href{https://arxiv.org/abs/astro-ph/0603494}{{\ttfamily
  astro-ph/0603494}}].

\bibitem{Avignone:2007fu}
F.~T. Avignone, III, S.~R. Elliott and J.~Engel, \emph{{Double Beta Decay,
  Majorana Neutrinos, and Neutrino Mass}},
  \href{https://doi.org/10.1103/RevModPhys.80.481}{\emph{Rev. Mod. Phys.}
  {\bfseries 80} (2008) 481--516},
  [\href{https://arxiv.org/abs/0708.1033}{{\ttfamily 0708.1033}}].

\bibitem{texono_2006}
{\scshape TEXONO} collaboration, H.~T. Wong et~al., \emph{{A Search of Neutrino
  Magnetic Moments with a High-Purity Germanium Detector at the Kuo-Sheng
  Nuclear Power Station}},
  \href{https://doi.org/10.1103/PhysRevD.75.012001}{\emph{Phys. Rev. D}
  {\bfseries 75} (2007) 012001},
  [\href{https://arxiv.org/abs/hep-ex/0605006}{{\ttfamily hep-ex/0605006}}].

\bibitem{beda_2013}
A.~G. Beda, V.~B. Brudanin, V.~G. Egorov, D.~V. Medvedev, V.~S. Pogosov, E.~A.
  Shevchik et~al., \emph{{Gemma experiment: The results of neutrino magnetic
  moment search}}, \href{https://doi.org/10.1134/S1547477113020027}{\emph{Phys.
  Part. Nucl. Lett.} {\bfseries 10} (2013) 139--143}.

\bibitem{borexino_2000}
{\scshape Borexino} collaboration, G.~Alimonti et~al., \emph{{Science and
  technology of BOREXINO: A Real time detector for low-energy solar
  neutrinos}},
  \href{https://doi.org/10.1016/S0927-6505(01)00110-4}{\emph{Astropart. Phys.}
  {\bfseries 16} (2002) 205--234},
  [\href{https://arxiv.org/abs/hep-ex/0012030}{{\ttfamily hep-ex/0012030}}].

\bibitem{borexino_2013}
{\scshape Borexino} collaboration, G.~Bellini et~al., \emph{{Final results of
  Borexino Phase-I on low energy solar neutrino spectroscopy}},
  \href{https://doi.org/10.1103/PhysRevD.89.112007}{\emph{Phys. Rev. D}
  {\bfseries 89} (2014) 112007},
  [\href{https://arxiv.org/abs/1308.0443}{{\ttfamily 1308.0443}}].

\bibitem{borexino2_2014}
{\scshape BOREXINO} collaboration, G.~Bellini et~al., \emph{{Neutrinos from the
  primary proton\textendash{}proton fusion process in the Sun}},
  \href{https://doi.org/10.1038/nature13702}{\emph{Nature} {\bfseries 512}
  (2014) 383--386}.

\bibitem{borexino_2017}
{\scshape Borexino} collaboration, M.~Agostini et~al., \emph{{First
  Simultaneous Precision Spectroscopy of $pp$, $^7$Be, and $pep$ Solar
  Neutrinos with Borexino Phase-II}},
  \href{https://doi.org/10.1103/PhysRevD.100.082004}{\emph{Phys. Rev. D}
  {\bfseries 100} (2019) 082004},
  [\href{https://arxiv.org/abs/1707.09279}{{\ttfamily 1707.09279}}].

\bibitem{coherent_2017}
{\scshape COHERENT} collaboration, D.~Akimov et~al., \emph{{Observation of
  Coherent Elastic Neutrino-Nucleus Scattering}},
  \href{https://doi.org/10.1126/science.aao0990}{\emph{Science} {\bfseries 357}
  (2017) 1123--1126}, [\href{https://arxiv.org/abs/1708.01294}{{\ttfamily
  1708.01294}}].

\bibitem{connie_2019}
{\scshape CONNIE} collaboration, A.~Aguilar-Arevalo et~al., \emph{{Search for
  light mediators in the low-energy data of the CONNIE reactor neutrino
  experiment}}, \href{https://doi.org/10.1007/JHEP04(2020)054}{\emph{JHEP}
  {\bfseries 04} (2020) 054},
  [\href{https://arxiv.org/abs/1910.04951}{{\ttfamily 1910.04951}}].

\bibitem{conus_2020}
{\scshape CONUS} collaboration, H.~Bonet et~al., \emph{{Constraints on elastic
  neutrino nucleus scattering in the fully coherent regime from the CONUS
  experiment}},
  \href{https://doi.org/10.1103/PhysRevLett.126.041804}{\emph{Phys. Rev. Lett.}
  {\bfseries 126} (2021) 041804},
  [\href{https://arxiv.org/abs/2011.00210}{{\ttfamily 2011.00210}}].

\bibitem{ccm_2021}
{\scshape CCM} collaboration, A.~A. Aguilar-Arevalo et~al., \emph{{First dark
  matter search results from Coherent CAPTAIN-Mills}},
  \href{https://doi.org/10.1103/PhysRevD.106.012001}{\emph{Phys. Rev. D}
  {\bfseries 106} (2022) 012001},
  [\href{https://arxiv.org/abs/2105.14020}{{\ttfamily 2105.14020}}].

\bibitem{colaresi_2022}
J.~Colaresi, J.~I. Collar, T.~W. Hossbach, C.~M. Lewis and K.~M. Yocum,
  \emph{{Measurement of Coherent Elastic Neutrino-Nucleus Scattering from
  Reactor Antineutrinos}},
  \href{https://doi.org/10.1103/PhysRevLett.129.211802}{\emph{Phys. Rev. Lett.}
  {\bfseries 129} (2022) 211802},
  [\href{https://arxiv.org/abs/2202.09672}{{\ttfamily 2202.09672}}].

\bibitem{pilar_2022}
P.~Coloma, I.~Esteban, M.~C. Gonzalez-Garcia, L.~Larizgoitia, F.~Monrabal and
  S.~Palomares-Ruiz, \emph{{Bounds on new physics with data of the Dresden-II
  reactor experiment and COHERENT}},
  \href{https://doi.org/10.1007/JHEP05(2022)037}{\emph{JHEP} {\bfseries 05}
  (2022) 037}, [\href{https://arxiv.org/abs/2202.10829}{{\ttfamily
  2202.10829}}].

\bibitem{xenon_1t}
{\scshape XENON} collaboration, E.~Aprile et~al., \emph{{First Dark Matter
  Search Results from the XENON1T Experiment}},
  \href{https://doi.org/10.1103/PhysRevLett.119.181301}{\emph{Phys. Rev. Lett.}
  {\bfseries 119} (2017) 181301},
  [\href{https://arxiv.org/abs/1705.06655}{{\ttfamily 1705.06655}}].

\bibitem{xenonnt_resolution_2020}
{\scshape XENON} collaboration, E.~Aprile et~al., \emph{{Excess electronic
  recoil events in XENON1T}},
  \href{https://doi.org/10.1103/PhysRevD.102.072004}{\emph{Phys. Rev. D}
  {\bfseries 102} (2020) 072004},
  [\href{https://arxiv.org/abs/2006.09721}{{\ttfamily 2006.09721}}].

\bibitem{lz_22}
{\scshape LZ} collaboration, J.~Aalbers et~al., \emph{{First Dark Matter Search
  Results from the LUX-ZEPLIN (LZ) Experiment}},
  \href{https://arxiv.org/abs/2207.03764}{{\ttfamily 2207.03764}}.

\bibitem{xenonnt_2022}
{\scshape XENON} collaboration, E.~Aprile et~al., \emph{{Search for New Physics
  in Electronic Recoil Data from XENONnT}},
  \href{https://doi.org/10.1103/PhysRevLett.129.161805}{\emph{Phys. Rev. Lett.}
  {\bfseries 129} (2022) 161805},
  [\href{https://arxiv.org/abs/2207.11330}{{\ttfamily 2207.11330}}].

\bibitem{cdex_2022}
{\scshape CDEX} collaboration, X.~P. Geng et~al., \emph{{Search for exotic
  neutrino interactions using solar neutrinos in the CDEX-10 experiment}},
  \href{https://arxiv.org/abs/2210.01604}{{\ttfamily 2210.01604}}.

\bibitem{montanino_2008}
D.~Montanino, M.~Picariello and J.~Pulido, \emph{{Probing neutrino magnetic
  moment and unparticle interactions with Borexino}},
  \href{https://doi.org/10.1103/PhysRevD.77.093011}{\emph{Phys. Rev. D}
  {\bfseries 77} (2008) 093011},
  [\href{https://arxiv.org/abs/0801.2643}{{\ttfamily 0801.2643}}].

\bibitem{harnik_2012}
R.~Harnik, J.~Kopp and P.~A.~N. Machado, \emph{{Exploring nu Signals in Dark
  Matter Detectors}},
  \href{https://doi.org/10.1088/1475-7516/2012/07/026}{\emph{JCAP} {\bfseries
  07} (2012) 026}, [\href{https://arxiv.org/abs/1202.6073}{{\ttfamily
  1202.6073}}].

\bibitem{agarwalla_2012}
S.~K. Agarwalla, F.~Lombardi and T.~Takeuchi, \emph{{Constraining Non-Standard
  Interactions of the Neutrino with Borexino}},
  \href{https://doi.org/10.1007/JHEP12(2012)079}{\emph{JHEP} {\bfseries 12}
  (2012) 079}, [\href{https://arxiv.org/abs/1207.3492}{{\ttfamily 1207.3492}}].

\bibitem{khan_2019}
A.~N. Khan, W.~Rodejohann and X.-J. Xu, \emph{{Borexino and general neutrino
  interactions}},
  \href{https://doi.org/10.1103/PhysRevD.101.055047}{\emph{Phys. Rev. D}
  {\bfseries 101} (2020) 055047},
  [\href{https://arxiv.org/abs/1906.12102}{{\ttfamily 1906.12102}}].

\bibitem{amaral_2020}
D.~W. P.~d. Amaral, D.~G. Cerdeno, P.~Foldenauer and E.~Reid, \emph{{Solar
  neutrino probes of the muon anomalous magnetic moment in the gauged $
  \mathrm{U}{(1)}_{L_{\mu }-{L}_{\tau }} $}},
  \href{https://doi.org/10.1007/JHEP12(2020)155}{\emph{JHEP} {\bfseries 12}
  (2020) 155}, [\href{https://arxiv.org/abs/2006.11225}{{\ttfamily
  2006.11225}}].

\bibitem{chen_2021}
Z.~Chen, T.~Li and J.~Liao, \emph{{Constraints on general neutrino interactions
  with exotic fermion from neutrino-electron scattering experiments}},
  \href{https://doi.org/10.1007/JHEP05(2021)131}{\emph{JHEP} {\bfseries 05}
  (2021) 131}, [\href{https://arxiv.org/abs/2102.09784}{{\ttfamily
  2102.09784}}].

\bibitem{Jho:2020sku}
Y.~Jho, J.-C. Park, S.~C. Park and P.-Y. Tseng, \emph{{Leptonic New Force and
  Cosmic-ray Boosted Dark Matter for the XENON1T Excess}},
  \href{https://doi.org/10.1016/j.physletb.2020.135863}{\emph{Phys. Lett. B}
  {\bfseries 811} (2020) 135863},
  [\href{https://arxiv.org/abs/2006.13910}{{\ttfamily 2006.13910}}].

\bibitem{Abdullah:2022zue}
M.~Abdullah et~al., \emph{{Coherent elastic neutrino-nucleus scattering:
  Terrestrial and astrophysical applications}},
  \href{https://arxiv.org/abs/2203.07361}{{\ttfamily 2203.07361}}.

\bibitem{coloma_2022}
P.~Coloma, M.~C. Gonzalez-Garcia, M.~Maltoni, J.~P. Pinheiro and S.~Urrea,
  \emph{{Constraining new physics with Borexino Phase-II spectral data}},
  \href{https://doi.org/10.1007/JHEP07(2022)138}{\emph{JHEP} {\bfseries 07}
  (2022) 138}, [\href{https://arxiv.org/abs/2204.03011}{{\ttfamily
  2204.03011}}].

\bibitem{khan_2022}
A.~N. Khan, \emph{{Light new physics and neutrino electromagnetic interactions
  in XENONnT}},
  \href{https://doi.org/10.1016/j.physletb.2022.137650}{\emph{Phys. Lett. B}
  {\bfseries 837} (2023) 137650},
  [\href{https://arxiv.org/abs/2208.02144}{{\ttfamily 2208.02144}}].

\bibitem{dutta_2022}
B.~Dutta, S.~Ghosh, T.~Li, A.~Thompson and A.~Verma, \emph{{Non-standard
  neutrino interactions in light mediator models at reactor experiments}},
  \href{https://doi.org/10.1007/JHEP03(2023)163}{\emph{JHEP} {\bfseries 03}
  (2023) 163}, [\href{https://arxiv.org/abs/2209.13566}{{\ttfamily
  2209.13566}}].

\bibitem{zeff_2022}
M.~Atzori~Corona, W.~M. Bonivento, M.~Cadeddu, N.~Cargioli and F.~Dordei,
  \emph{{New constraint on neutrino magnetic moment and neutrino millicharge
  from LUX-ZEPLIN dark matter search results}},
  \href{https://doi.org/10.1103/PhysRevD.107.053001}{\emph{Phys. Rev. D}
  {\bfseries 107} (2023) 053001},
  [\href{https://arxiv.org/abs/2207.05036}{{\ttfamily 2207.05036}}].

\bibitem{rahul_2022}
S.~K. A., A.~Majumdar, D.~K. Papoulias, H.~Prajapati and R.~Srivastava,
  \emph{{Implications of first LZ and XENONnT results: A comparative study of
  neutrino properties and light mediators}},
  \href{https://doi.org/10.1016/j.physletb.2023.137742}{\emph{Phys. Lett. B}
  {\bfseries 839} (2023) 137742},
  [\href{https://arxiv.org/abs/2208.06415}{{\ttfamily 2208.06415}}].

\bibitem{miranda_2015}
O.~G. Miranda and H.~Nunokawa, \emph{{Non standard neutrino interactions:
  current status and future prospects}},
  \href{https://doi.org/10.1088/1367-2630/17/9/095002}{\emph{New J. Phys.}
  {\bfseries 17} (2015) 095002},
  [\href{https://arxiv.org/abs/1505.06254}{{\ttfamily 1505.06254}}].

\bibitem{bilmis_2015}
S.~Bilmis, I.~Turan, T.~M. Aliev, M.~Deniz, L.~Singh and H.~T. Wong,
  \emph{{Constraints on Dark Photon from Neutrino-Electron Scattering
  Experiments}}, \href{https://doi.org/10.1103/PhysRevD.92.033009}{\emph{Phys.
  Rev. D} {\bfseries 92} (2015) 033009},
  [\href{https://arxiv.org/abs/1502.07763}{{\ttfamily 1502.07763}}].

\bibitem{bhupal_2021}
P.~S.~B. Dev, D.~Kim, K.~Sinha and Y.~Zhang, \emph{{New interference effects
  from light gauge bosons in neutrino-electron scattering}},
  \href{https://doi.org/10.1103/PhysRevD.104.075001}{\emph{Phys. Rev. D}
  {\bfseries 104} (2021) 075001},
  [\href{https://arxiv.org/abs/2105.09309}{{\ttfamily 2105.09309}}].

\bibitem{borexino_2008}
{\scshape Borexino} collaboration, C.~Arpesella et~al., \emph{{Direct
  Measurement of the Be-7 Solar Neutrino Flux with 192 Days of Borexino Data}},
  \href{https://doi.org/10.1103/PhysRevLett.101.091302}{\emph{Phys. Rev. Lett.}
  {\bfseries 101} (2008) 091302},
  [\href{https://arxiv.org/abs/0805.3843}{{\ttfamily 0805.3843}}].

\bibitem{end_point}
D.~Baxter et~al., \emph{{Recommended conventions for reporting results from
  direct dark matter searches}},
  \href{https://doi.org/10.1140/epjc/s10052-021-09655-y}{\emph{Eur. Phys. J. C}
  {\bfseries 81} (2021) 907},
  [\href{https://arxiv.org/abs/2105.00599}{{\ttfamily 2105.00599}}].

\bibitem{parke_1986}
S.~J. Parke, \emph{{Nonadiabatic Level Crossing in Resonant Neutrino
  Oscillations}},
  \href{https://doi.org/10.1103/PhysRevLett.57.1275}{\emph{Phys. Rev. Lett.}
  {\bfseries 57} (1986) 1275--1278},
  [\href{https://arxiv.org/abs/2212.06978}{{\ttfamily 2212.06978}}].

\bibitem{fit_oscillation_2020}
P.~F. de~Salas, D.~V. Forero, S.~Gariazzo, P.~Mart\'\i{}nez-Mirav\'e, O.~Mena,
  C.~A. Ternes et~al., \emph{{2020 global reassessment of the neutrino
  oscillation picture}},
  \href{https://doi.org/10.1007/JHEP02(2021)071}{\emph{JHEP} {\bfseries 02}
  (2021) 071}, [\href{https://arxiv.org/abs/2006.11237}{{\ttfamily
  2006.11237}}].

\bibitem{rrpa_2016}
J.-W. Chen, H.-C. Chi, C.~P. Liu and C.-P. Wu, \emph{{Low-energy electronic
  recoil in xenon detectors by solar neutrinos}},
  \href{https://doi.org/10.1016/j.physletb.2017.10.029}{\emph{Phys. Lett. B}
  {\bfseries 774} (2017) 656--661},
  [\href{https://arxiv.org/abs/1610.04177}{{\ttfamily 1610.04177}}].

\bibitem{flux_uncertainity}
A.~Serenelli, \emph{{Alive and well: a short review about standard solar
  models}}, \href{https://doi.org/10.1140/epja/i2016-16078-1}{\emph{Eur. Phys.
  J. A} {\bfseries 52} (2016) 78},
  [\href{https://arxiv.org/abs/1601.07179}{{\ttfamily 1601.07179}}].

\bibitem{juno_2015}
{\scshape JUNO} collaboration, F.~An et~al., \emph{{Neutrino Physics with
  JUNO}}, \href{https://doi.org/10.1088/0954-3899/43/3/030401}{\emph{J. Phys.
  G} {\bfseries 43} (2016) 030401},
  [\href{https://arxiv.org/abs/1507.05613}{{\ttfamily 1507.05613}}].

\bibitem{Seo:2020dtx}
S.~H. Seo and Y.~D. Kim, \emph{{Dark Photon Search at Yemilab, Korea}},
  \href{https://doi.org/10.1007/JHEP04(2021)135}{\emph{JHEP} {\bfseries 04}
  (2021) 135}, [\href{https://arxiv.org/abs/2009.11155}{{\ttfamily
  2009.11155}}].

\bibitem{lsc_2022}
J.~R. Alonso et~al., \emph{{IsoDAR@Yemilab: A report on the technology,
  capabilities, and deployment}},
  \href{https://doi.org/10.1088/1748-0221/17/09/P09042}{\emph{JINST} {\bfseries
  17} (2022) P09042}, [\href{https://arxiv.org/abs/2201.10040}{{\ttfamily
  2201.10040}}].

\end{thebibliography}
\end{document}